\documentclass[a4paper, 10pt, nofootinbib]{revtex4}
\usepackage[top=20mm,bottom=20mm,left=15mm,right=15mm]{geometry}
\usepackage{amsmath, amssymb, amsfonts, mathtools, comment, cases, braket, amsthm, bm, mathdots, cancel, bm, float}

\numberwithin{equation}{section} 

\usepackage{hyperref}
\hypersetup{
 setpagesize=false,
 bookmarksnumbered=true,%
 bookmarksopen=true,%
 colorlinks=true,%
 linkcolor=blue,
 citecolor=red,
}

\newcommand{\be}{\begin{equation}}
\newcommand{\ee}{\end{equation}}

\newcommand{\myalign}[1]{\begin{align}#1\end{align}}

\newcommand{\Mpl}{M_{\rm pl}}

\begin{document}

\title{Hairy black holes in extended Einstein-Maxwell-scalar theories with magnetic charge and kinetic couplings}
\author{Kitaro Taniguchi, Shinta Takagishi, and Ryotaro Kase}
\affiliation{
Department of Physics, Faculty of Science, 
Tokyo University of Science, 1-3, Kagurazaka,
Shinjuku-ku, Tokyo 162-8601, Japan
}
\begin{abstract}
We study static and spherically symmetric black hole (BH) solutions in extended Einstein-Maxwell-scalar theories, 
which are classified in a subclass of the $U(1)$ gauge-invariant scalar-vector-tensor theories.
The scalar field is coupled to the vector field, which has electric and magnetic charges.
We investigate modifications to the Reissner-Nordstr\"{o}m solution focusing on 
the three types of scalar-vector interactions, including derivative couplings.
We solve the field equations analytically in two asymptotic regions
which are the vicinity of the BH horizon and the spatial infinity, and 
clarify the condition for the existence of scalar hair.
To understand the behaviors of solutions in intermediate scales,
the field equations are integrated numerically for concrete models 
with different types of couplings.
We find new hairy BH solutions with scalar hair 
in the presence of magnetic charge and kinetic coupling.
The magnetic charge plays an important role in distinguishing hairy BH solutions 
originated from three types of different interactions at a large coupling limit.
\end{abstract}

\date{\today}

\maketitle
%
%
%
\section{Introduction}
Experimental tests of General Relativity (GR) is well performed in the solar system \cite{Will:2014kxa}.
As a result, it is known that the weak gravitational field can be described by GR in high precision.
On the other hand, the accuracy of GR in the strong field regime has been not yet clarified well enough.
In recent years, the observations of the nonlinear regime of gravity have represented a significant development.
For example, the direct detection of gravitational waves 
from compact binary coalescences by LIGO-Virgo-KAGRA Collaborations
\cite{LIGOScientific:2016aoc, LIGOScientific:2016sjg, LIGOScientific:2017bnn, LIGOScientific:2017vox, LIGOScientific:2017ycc, LIGOScientific:2017vwq, LIGOScientific:2018mvr, LIGOScientific:2020stg, LIGOScientific:2020aai, LIGOScientific:2020zkf, LIGOScientific:2021qlt, KAGRA:2021vkt}
and the imaging of black hole (BH) shadow by the Event Horizon Telescope
\cite{EventHorizonTelescope:2019dse, EventHorizonTelescope:2022wkp} 
opened up a new window for exploring physics around BHs.
The action including Maxwell electromagnetic field besides GR is called Einstein-Maxwell theory.
The authors of Ref.~\cite{Colleaux:2023cqu} have classified the generalized higher-order Einstein-Maxwell Lagrangians, 
which consist of terms linear in the Riemann tensor and quadratic in derivatives of the field strength tensor.
In the four dimensional Einstein-Maxwell system without matter, asymptotically flat and stationary BH solutions
are described by only three parameters, i.e., mass, electric charge, and angular momentum
\cite{Israel:1967wq, Carter:1971zc, Ruffini:1971bza, Hawking:1971vc}.
Hence, the Kerr-Newman BH \cite{Newman:1965my} is the unique solution in such a system.
This statement is known as the ``no-hair theorem.''
It is natural to wonder whether this theorem still holds even in the presence of a new degree of freedom. 
If a canonical scalar field minimally coupled to gravity is introduced to the Einstein-Maxwell system, 
then, the profile of scalar field settles down to be trivial
\cite{Chase:1970omy, Bekenstein:1971hc, Bekenstein:1996pn}
, and the spacetime is still described by the Kerr-Newman solutions.
This fact shows the absence of a ``scalar hair'' due to the validity of the no-hair theorem 
even in the presence of the canonical scalar field.
Constructing BH solutions with the ``scalar hair'' in the context of GR is not a simple problem
\cite{Chew:2022enh, Chew:2023olq, Ghosh:2023kge, Chew:2024rin}
(see Ref.~\cite{Herdeiro:2015waa} for review). 
However, if a scalar field couples to the Maxwell field, it can take a nontrivial profile
\cite{Gibbons:1982ih, Gibbons:1987ps, Garfinkle:1990qj, Campbell:1991rz, Lee:1991jw, Reuter:1991cb, Monni:1995vu, Stefanov:2007qw, Stefanov:2007bn, Stefanov:2007eq, Sheykhi:2014gia, Sheykhi:2014ipa, Sheykhi:2015ira, Fan:2015oca, Dehghani:2018eps, Herdeiro:2018wub, Myung:2018vug, Herdeiro:2019oqp, Myung:2019oua, Konoplya:2019goy, Hod:2019ulh, Dehghani:2019cuf, Filippini:2019cqk, Boskovic:2018lkj, Fernandes:2019kmh, Astefanesei:2019pfq, Herdeiro:2019tmb, Brihaye:2019gla, Fernandes:2020gay, Mahapatra:2020wym, Priyadarshinee:2021rch, Priyadarshinee:2023cmi, Promsiri:2023yda, Belkhadria:2023ooc}.
Such a coupling between scalar and vector fields arises in several theories, e.g., 
effective Lagrangians of axion \cite{Peccei:1977hh, Georgi:1986df} which is a strong candidate of dark matter 
and compactified higher-dimensional theories \cite{Gibbons:1982ih, Gibbons:1987ps, Garfinkle:1990qj}.
\par 
The interactions between scalar and vector fields can be generalized requiring 
the equations of motion being up to second order in terms of derivatives for 
the absence of Ostrogradsky instabilities
\cite{Ostrogradsky:1850fid, Woodard:2015zca}.
Indeed, the application of this method to a single scalar degree of freedom nonminimally 
coupled to gravity leads the Horndeski theory, which is the most general scalar-tensor theory 
with second-order equations of motion 
\cite{Horndeski:1974wa, Deffayet:2011gz, Kobayashi:2011nu}\footnote{
Horndeski theories can be further generalized so as to allow the existence of higher order derivatives 
while keeping the number of propagating degree of freedom by virtue of degeneracy conditions. 
This class of theories is called Degenerate Higher-Order Scalar-Tensor (DHOST) theories 
\cite{Langlois:2015cwa}.}. 
For the massive vector degree of freedom, the same approach results in the generalized Proca theory
\cite{Heisenberg:2014rta, Tasinato:2014eka, Tasinato:2014mia}.
The Horndeski and the generalized Proca theories can be unified 
as the scalar-vector-tensor (SVT) theories in association with 
new interactions between the scalar field and the vector field
\cite{Heisenberg:2018acv}. If the vector field satisfies 
the $U(1)$ gauge symmetry especially, it reduces to the so-called 
$U(1)$ gauge-invariant SVT theories in which the Einstein-Maxwell-scalar theory 
is classified as a subclass.
\par 
The interactions between scalar field and vector field in the $U(1)$ gauge-invariant SVT theories 
are described by three Lagrangians denoted as $\mathcal{L}_2, \mathcal{L}_3$, and $\mathcal{L}_4$. 
The scalar or axionic-type couplings to the vector field discussed in Refs. 
\cite{Gibbons:1987ps, Garfinkle:1990qj, Campbell:1991rz, Lee:1991jw, Reuter:1991cb, Sheykhi:2014gia, Fan:2015oca, Myung:2018vug, Boskovic:2018lkj, Myung:2019oua, Dehghani:2019cuf, Filippini:2019cqk, Fernandes:2019kmh, Astefanesei:2019pfq, Herdeiro:2019tmb, Fernandes:2020gay} 
correspond to particular cases of the $\mathcal{L}_2$ interaction\footnote{
The existence of hairy BH solutions arising from the interactions $\mathcal{L}_3$ 
and $\mathcal{L}_4$ are discovered in Ref.~\cite{Heisenberg:2018vti}, and the 
stability of these solutions against odd-parity perturbations are studied in 
Ref.~\cite{Heisenberg:2018mgr}.}. 
In these works, the scalar field $\phi$ couples to the vector field through the scalar type interaction 
of the form $g_1(\phi)F_{\mu \nu}F^{\mu \nu}$ or the axionic-type interaction of the form 
$g_2(\phi)F_{\mu\nu} \tilde{F}^{\mu\nu}$ where 
$F_{\mu \nu}$ is the strength tensor of vector field and $\tilde{F}_{\mu\nu}$ is its dual, 
$g_1$ and $g_2$ are functions of $\phi$. 
On the other hand, in the range of the $U(1)$ gauge-invariant SVT theories, one can consider 
other types of interactions. 
We can take account of the field derivatives $\nabla_{\alpha}\phi$, instead of 
the field $\phi$ itself, coupled to the vector field as 
$\nabla_{\alpha}\phi \nabla^{\alpha}\phi\, F_{\mu\nu} F^{\mu\nu}$ and 
$\nabla_{\mu}\phi \nabla^{\nu}\phi\, F^{\mu\alpha} F_{\nu\alpha}$
where $\nabla_{\mu}$ is the covariant derivative operator.
If we demand the $U(1)$ gauge field to be static and spherically symmetric
at the time of configuration, however, the latter interaction, 
$\nabla_{\mu}\phi \nabla^{\nu}\phi\, F^{\mu\alpha} F_{\nu\alpha}$, reduces to 
be in proportion to the former interaction, 
$\nabla_{\alpha}\phi \nabla^{\alpha}\phi\, F_{\mu\nu} F^{\mu\nu}$.
This shows that the two types of interactions are degenerated on the configuration 
and cannot be distinguished.
Moreover, we cannot examine the effect of a scalar product $F_{\mu\nu} \tilde{F}^{\mu\nu}$ 
at the background since it identically vanishes.
\par 
One way to resolve the aforementioned degeneracy between 
$\nabla_{\alpha}\phi \nabla^{\alpha}\phi\, F_{\mu\nu} F^{\mu\nu}$ and 
$\nabla_{\mu}\phi \nabla^{\nu}\phi\, F^{\mu\alpha} F_{\nu\alpha}$ 
on the static and spherically symmetric background is to introduce the magnetic charge. 
Although the existence of magnetic charge breaks the spherical symmetry of the vector field, 
it does not ruin the same symmetry at the level of the background equations. 
In this case, it is known that the term $F_{\mu\nu}\tilde{F}^{\mu\nu}$ survives at the background 
and realizes the hairy BH solution with a scalar hair through the axionic-type interaction 
of the form $g_2(\phi)F_{\mu\nu}\tilde{F}^{\mu\nu}$ \cite{Lee:1991jw, Fernandes:2019kmh}.
Several spherically symmetric solutions in the presence of a magnetic monopole (e.g., Dirac monopole \cite{Dirac:1931kp}) 
are also discussed in Refs.~\cite{Gibbons:1990um, Ortiz:1991eu, Lee:1991vy, Lee:1991qs, Breitenlohner:1991aa, Bronnikov:2000vy, Ayon-Beato:2000mjt, Fernandes:2019kmh, Nomura:2020tpc, Maldacena:2020skw}. 
\par 
In this paper, we study BH solutions in extended Einstein-Maxwell-scalar theories 
in the presence of magnetic charge on the static and spherically symmetric spacetime. 
The existence of magnetic charge enables us to distinguish the two types of derivative 
interactions $\nabla_{\mu}\phi \nabla^{\nu}\phi\, F^{\mu\alpha} F_{\nu\alpha}$ and 
$\nabla_{\alpha}\phi \nabla^{\alpha}\phi\, F^{\mu\nu} F_{\mu\nu}$. 
We show that the former interaction can give rise to a new type of hairy BH solutions. 
Moreover, we also investigate the possible modifications to the Reissner-Nordstr\"{o}m (RN) solutions 
by including the latter type of interactions by extending the scalar and axionic-type couplings as 
$g_1(\phi ,X)F_{\mu \nu}F^{\mu \nu}$ and $g_2(\phi,X)F_{\mu\nu} \tilde{F}^{\mu\nu}$ 
where $X=-\nabla_{\alpha}\phi \nabla^{\alpha}\phi/2$ is the kinetic  term of scalar field. 
\par 
This paper is organized as follows.
In Sec.~\ref{sec2}, we present the Lagrangian describing scalar-vector couplings 
which corresponds to the quadratic term $\mathcal{L}_2$ in the $U(1)$ gauge-invariant 
SVT theories, and focus on the three types of scalar-vector coupling. 
The vector potential is set to have a magnetic charge.
We derive the field equations on a static and spherically symmetric background.
In Sec.~\ref{sec3}, we obtain analytic solutions by expanding the field equations 
in the two asymptotic regions, i.e., in the vicinity of the BH horizon and at the spatial infinity. 
We investigate the condition for the presence of scalar hair by using these solutions. 
In Sec.~\ref{sec4}, we apply our results derived in the previous section to the concrete models 
, and confirm the existence of scalar hair in the intermediate regions by numerically integrating 
the background equations. We then discuss the possible parameter region to realize 
the BH solutions as well as the difference of each BH solution in the large coupling limit. 
The deviations from the RN solution, especially, are investigated, focusing on the effects of derivative couplings.  
The Sec.~\ref{sec5} is devoted to conclusions.
We use the geometrized units, where the speed of light and the gravitational constant are equal to $1$.
%
%
%
\section{$U(1)$ gauge-invariant SVT theories and field equations}
\label{sec2}
%
We consider a subclass of the $U(1)$ gauge-invariant scalar-vector-tensor (SVT) theories described by the action  \cite{Heisenberg:2018acv},
\be
	S = \int {\rm d}^4 x \sqrt{-g}
			\left[
				{\Mpl^2 \over 2} R + f_2 (\phi, X, F, \tilde{F}, Y)
			\right]\,,
			\label{eq:act_SVT}
\ee
where $g$ is a determinant of the metric tensor $g_{\mu \nu}$,
and the term proportional to the Ricci scalar $R$ is the Einstein-Hilbert term
with the reduced Planck mass $\Mpl=1/\sqrt{8\pi G}$.
The function $f_2$ corresponding to the quadratic Lagrangian $\mathcal{L}_2$ in the 
$U(1)$ gauge-invariant SVT theories depends on scalar field $\phi$ and following quantities,
\myalign{
	&X = - {1\over2} \nabla_{\mu} \phi \nabla^{\mu} \phi\,, \qquad
	F = - {1\over4} F_{\mu \nu} F^{\mu \nu}\,, \qquad
	\tilde{F} = - {1\over4} F_{\mu \nu} \tilde{F}^{\mu \nu}\,,\label{defX}
\\
	&F_{\mu \nu} = \nabla_{\mu} A_{\nu} - \nabla_{\nu} A_{\mu}\,, \qquad
	\tilde{F}^{\mu \nu} 
	= {1\over2} \mathcal{E}^{\mu\nu\alpha\beta} F_{\alpha\beta}\,, \qquad
	Y = \nabla_{\mu} \phi \nabla^{\nu} \phi F^{\mu \alpha} F_{\nu \alpha}\,,\label{defY}
}
where $A_{\mu}$ is the vector field, $\nabla_{\mu}$ is the covariant derivative operator, and
$\mathcal{E}^{\mu \nu \alpha \beta}$ is the anti-symmetric Levi-Civita tensor 
satisfying the normalization $\mathcal{E}^{\mu \nu \alpha \beta} \mathcal{E}_{\mu \nu \alpha \beta} = +4!$.
\par
We focus on a static and spherically symmetric spacetime with the line-element,
\be
	{\rm d}s^2 = -f(r) {\rm d}t^2 + h^{-1}(r) {\rm d}r^2 + r^2 {\rm d}\Omega^2\,, 
	\qquad {\rm d}\Omega^2 = {\rm d}\theta^2 + \sin^2 \theta {\rm d}\varphi^2\,,
	\label{eq:line-element}
\ee
where $t$ is the time coordinate, $r$ stands for the radial coordinate,
$\theta$ is polar angle, and $\varphi$ is azimuthal angle.
The two functions $f$ and $h$ depend on only $r$.
We define the event horizon radius $r = r_h$ where $f(r_h)$ and $h(r_h)$ vanish simultaneously.
Furthermore, we assume that $f(r)$ and $h(r)$ remain positive outside the horizon ($r > r_h$).
According to the symmetry of background metric, we adopt the following field ansatz
\be
	\phi = \phi(r)\,,\qquad
	A_{\mu} = (V(r), 0, 0, - P \cos \theta),
	\label{eq:ansatz}
\ee
where $\phi$, $V$ are functions of the radial coordinate $r$, 
and $P$ corresponds to the monopolar magnetic charge.
From Eqs.~\eqref{eq:line-element} and \eqref{eq:ansatz}, the scalar quantities 
$X$, $F$, $\tilde{F},$ and $Y$, defined in Eqs.~\eqref{defX}-\eqref{defY}, are 
evaluated as 
\be
	X = - \frac{h \phi'^2}{2}\,,\qquad
	F = \frac{h V'^2}{2f} - \frac{P^2}{2r^4}\,,\qquad
	\tilde{F} = \frac{P V'}{r^2} \sqrt{\frac{h}{f}}\,,\qquad
	Y = - \frac{h^2 \phi'^2 V'^2}{f}.
	\label{eq:bg_quantities}
\ee
In the absence of magnetic charge $P$, the quantity $\tilde{F}$ identically vanishes and
the derivative interaction $Y$ reduces to $Y = 4 X F$~\cite{Heisenberg:2017hwb, Kase:2018voo}. 
This means that we can omit the dependence of these quantities in the function $f_2$ in Eq.~\eqref{eq:act_SVT} 
without loss of generality at the background level. 
On the other hand, if the nonzero magnetic charge exists, the situation changes significantly. 
The quantity $\tilde{F}$ survives and $Y$ becomes independent of $F$. 
So far, the hairy BH solutions arising from the scalar type interaction $g_1(\phi)F_{\mu\nu}F^{\mu\nu}$ 
and the axionic-type interaction $g_2(\phi)F_{\mu\nu}\tilde{F}^{\mu\nu}$ have been broadly studied in 
Refs.~\cite{Gibbons:1987ps, Garfinkle:1990qj, Lee:1991jw, Boskovic:2018lkj, Astefanesei:2019pfq, Fernandes:2019kmh}. 
In this paper, we focus on the effect of derivative interactions on the hairy BH solutions 
by taking into account the derivative coupling $Y$ as a key ingredient 
as well as the $X$-dependence in the scalar and axionic-type interactions. 
For this purpose, we consider extended Einstein-Maxwell-scalar theories described by  
\be
	S = \int {\rm d}^4 x \sqrt{-g}
	\left[
		\frac{\Mpl^2}{2} R + F + X + g_1(\phi,X) F + g_2(\phi,X) \tilde{F} 
			+ \bar{g}_3(\phi,X) Y
	\right]\,,
	\label{eq:exEMS_action}
\ee
which belongs to the subclass of theories given in Eq.~\eqref{eq:act_SVT}. 
Here, $g_1, g_2$, and $\bar{g}_3$ are arbitrary functions of the scalar field $\phi$ and 
its kinetic term $X$. 
In the following sections, we will search for nontrivial BH solutions such that $f$, $h$, $\phi$, and 
$V$ are regular at the horizon. Assuming that the above Lagrangian is also regular at the horizon, 
$g_1$ and $g_2$ need to be functions that have zero or positive powers of $X$ since this quantity vanishes 
at the horizon (see Eq.~(\ref{eq:bg_quantities})). On the other hand, the situation is different for $g_3$. 
As can be seen from Eq.~(\ref{eq:bg_quantities}), the quantity $Y$ is proportional to $X$ in the form 
$Y=(2hV'^2/f)X$. This shows that, even for $\bar{g}_3 \propto X^{-1}$, the combination $\bar{g}_3Y$ is 
regular at the horizon. In order to make this point clear, we normalize $\bar{g}_3$ as 
\be
\bar{g}_3(\phi,X)=\frac{g_3(\phi,X)}{4X}\,, 
\label{eq:g3_normalize}
\ee
without loss of generality. If we demand $g_3$ has zero or positive power of $X$, the Lagrangian 
$\sqrt{-g}\,g_3Y/(4X)$ can be regular at the horizon. 
\par 
Variation of the action (\ref{eq:exEMS_action}) with respect to $f, h, \phi,$ and $V$ 
gives the equations of motion,
\myalign{
	2\Mpl^2 r f h' &= 2\Mpl^2 f (1-h)
										 - r^2 h \left[ f \phi'{}^2 + (1+g_1+g_3) V'{}^2 \right]
										 - {P^2 f (1+g_1) \over r^2}\,,
										 \label{eq:BG}
	\\
	2\Mpl^2 r h f' &= 2\Mpl^2 f (1-h)
										 + r^2 h \left[ f \phi'{}^2 - (1+g_1+g_3) V'{}^2 \right]
										 - {P^2 f (1+g_1) \over r^2} \nonumber \\
									 &\hspace{10pt} + \left[
										 	 	 \left(r^2 h (g_{1,X}+g_{3,X}) V'
												 + 2P \sqrt{fh}\, g_{2,X}\right) V'
												 - {P^2 f g_{1,X} \over r^2}
										 	 \right] h \phi'{}^2\,,
	\\
	J_{\phi}' &= \mathcal{P}_{\phi}\,, \label{eq:BG_phi}
	\\
	J_{A}' &= 0\,, \label{eq:BG_end}
}
respectively, where
\myalign{
	J_{\phi} &= -\left[
								\frac{r^2}{2} \sqrt{\frac{h}{f}}\,\, (g_{1,X} + g_{3,X})\, V'^2
								+ g_{2,X} P\,V'
								+ \sqrt{\frac{f}{h}} \left( r^2 - \frac{g_{1,X} P^2}{2 r^2} \right)
							\right] h\, \phi'\,, \label{defJphi}
	\\
	\mathcal{P}_{\phi}&=\frac{r^2}{2}\sqrt{\frac{h}{f}}\,\,(g_{1,\phi} + g_{3,\phi})\,V'^2
							+ g_{2,\phi}\,P\,V'
							- \sqrt{\frac{f}{h}}\,\,\frac{g_{1,\phi}\,P^2}{2r^2}\,,
	\\
	J_{A} &= r^2\sqrt{\frac{h}{f}}\,\,(1 + g_1 + g_3)\,V' + g_2\,P\,.\label{defJA}
}
Here, the scalar quantity that follows comma in subscripts represents a partial derivative
with respect to the quantity, e.g., $g_{i, \phi} = \partial g_i(\phi, X) / \partial \phi$ and
$g_{i, \phi \phi} = \partial^2 g_i(\phi, X) / \partial \phi^2$ (with $i \in \{1,2,3\}$),
whereas the prime denotes the derivatives with respect to $r$,
e.g., $\xi' = {\rm d}\xi / {\rm d}r$ (with $\xi \in \{f, h, V, \phi\}$).
From Eq.~(\ref{eq:BG_end}), we read $J_A = {\rm const.}$
This conservation of the current $J_A$ appears as a result of the $U(1)$ gauge symmetry.
%
%
%
\section{Asymptotic solutions}
\label{sec3}
We derive analytic solutions to the background equations (\ref{eq:BG})-(\ref{eq:BG_end}) in two asymptotic regions.
The metric components and field variables are required to be regular around the horizon.
At large distances from the origin of the BH, we demand the solutions to be asymptotically flat.
%
\subsection{Solutions around the horizon}
\label{sec3a}
To derive analytic solutions to the field equations (\ref{eq:BG})-(\ref{eq:BG_end}) 
in the vicinity of the BH horizon $r_h$, we expand $f, h, V, \phi$ with respect to $r-r_h$. 
Requiring the regularity of these variables as well as the vanishing $f$ and $h$ at the horizon, 
they can be expanded in the forms 
\be
	f = \sum_{i=1}^{\infty} f_i (r - r_h)^i\,,\qquad
	h = \sum_{i=1}^{\infty} h_i (r - r_h)^i\,,\qquad
	V = V_0 + \sum_{i=1}^{\infty} V_i (r - r_h)^i\,,\qquad
	\phi = \phi_0 + \sum_{i=1}^{\infty} \phi_i (r - r_h)^i,
	\label{eq:expansion_horizon}
\ee
where $f_i$, $h_i$, $V_0$, $V_i$, $\phi_0$, $\phi_i$ are constants. 
Hereafter, the constant $V_0$ will be omitted by setting $V_0 = 0$ without loss of generality,
since $V$ always appears with a derivative in the field equations by virtue of the $U(1)$ gauge symmetry.
Similar to the regularity in Eq.~(\ref{eq:expansion_horizon}), 
we assume that the coupling functions are regular at the horizon. Since the kinetic term $X$ is expanded 
around the horizon as $X=-(h_1\phi_1^2/2)(r-r_h)+{\cal O}((r-r_h)^2)$,  
the expansion of coupling $g_i(\phi,X)$ with $i=1,2,3$ can be expressed as 
$g_i(\phi,X)=g_i(\phi_0,0)+[g_{i,\phi}(\phi_0,0)-g_{i,X}(\phi_0,0)h_1\phi_1^2/2](r-r_h)+{\cal O}((r-r_h)^2)$. 
We denote $g_i(\phi_0,0)$, $g_{i,\phi}(\phi_0,0)$, $g_{i,X}(\phi_0,0)$ as $g_i$, $g_{i,\phi}$, $g_{i,X}$ 
for simplicity in the following discussion.
\par
We treat the effect of the couplings as corrections to 
the RN metric with a magnetic charge $P$ given by 
\be
	f_{\rm RN} = h_{\rm RN} = \left( 1 - \frac{r_h}{r} \right) \left( 1 - \mu \frac{r_h}{r} \right)
		= 1 - \frac{2M}{r} + \frac{Q^2 + P^2}{2 \Mpl^2 r^2}, \label{fRN}
\ee
where $r_h$ corresponds to the outer horizon and a dimensionless constant $\mu$ 
characterizes the inner horizon $\tilde{r}_h = \mu r_h$ within the range of $0 \leq \mu \leq 1$.
From Eq.~\eqref{fRN}, the constant $\mu$ satisfies $\mu = 2M / r_h - 1$, so that 
there is a relation $Q^2 + P^2 = 2 r_h (2 M - r_h) \Mpl^2$ among horizons and charges.
We substitute Eq.~\eqref{eq:expansion_horizon} into Eqs.~(\ref{eq:BG})-(\ref{eq:BG_end}) 
and expand them around $r\sim r_h$. 
Then, we iteratively solve the resultant equations order by order in terms of $r-r_h$
so as to derive the coefficients 
$f_i$, $h_i$, $V_i$, $\phi_i$. 
To clarify the leading contributions of the couplings to the modification of the RN solution, 
the iteration has been continued until the $g_1$, $g_2$, $g_3$ appear in all the coefficients 
$f_i$, $h_i$, $V_i$, $\phi_i$. 
Assuming that the quantities $f_1, h_1$ are positive since we are interested in the BH solutions 
outside the horizon, the coefficients are given by,
\myalign{
	f_1 &= h_1 = \frac{1 - \mu}{r_h}\,,\qquad \label{eq:sol_horizon}
	V_1 =  {1 \over r_h^2} 
				 {\sqrt{\gamma_{g_1 g_3} \left[2 \Mpl^2 r_h^2 \mu - (1 + g_1) P^2\right] } \over \gamma_{g_1 g_3}}\,, \\
	\phi_1 &= - {r_h \over 1 - \mu} \left({\Phi_{1,\phi} \over \Phi_{1,X} - 2 r_h^4}\right)\,,
		\label{eq:sol_horizon_phi1}
}
at the linear order, and
\myalign{
	f_2 &= {2\mu - 1 \over r_h^2} + \left({\Phi_{1,\phi} \over 8 \Mpl^2 r_h^3}\right) \phi_1
		\label{eq:sol_horizon_f2}
	\,,\qquad
	h_2 =  {2\mu - 1 \over r_h^2} - \left({3\, \Phi_{1,\phi} \over 8 \Mpl^2 r_h^3}\right) \phi_1
	\,, \\
	V_2 &= - {V_1\over r_h} 
				 + {1 \over 2r_h^2}
				 \left(
				   {V_1 \Phi_{1,\phi} \over 4 (1 - \mu) \Mpl^2 } - {\Phi_{2,\phi} \over \gamma_{g_1 g_3}}
				 \right) \phi_1
				 + \left({(1 - \mu)\, \Phi_{2,X} \over 4 r_h^3\, \gamma_{g_1 g_3}}\right) \phi_1^2
	\,, \\
	\phi_2 &= \left({3\mu - 1 \over 4 r_h (1 - \mu)}\right) \phi_1
							+ \left(
									{\Phi_{1,\phi} \over 8 (1 - \mu) \Mpl^2 r_h^2}
									+{(9\mu-5) \Phi_{1,X}+r_h^2 \left[\Phi_{1,\phi\phi}-2(\mu+3)r_h^2\right] 
										\over 
										4r_h^2  \Phi_{1,\phi}}
								\right.
						\\
						&\hspace{20pt}
								\left.
									- {3 \Phi_{1,\phi X} \over 8 \left(\Phi_{1,X}-2r_h^4\right)}
									+ {\Phi_{2,\phi} \over 2\gamma_{g_1g_3}} \left[
																				{\Phi_{2,\phi} \over \Phi_{1,\phi}}
																				 - {3\Phi_{2,X} \over 2(\Phi_{1,X}-2r_h^4)}
																			\right]
								\right) \phi_1^2
							+ \left(
									{(1-\mu) \left(\gamma_{g_1 g_3}\Phi_{1,XX}+2\Phi_{2,X}^2 \right) 
										\over 
									4r_h\gamma_{g_1 g_3} \left(\Phi_{1,X}-2r_h^4\right)}
								\right) \phi_1^3\,,
				\label{eq:sol_horizon_end}
}
at the second order, where
\myalign{
	\Phi_1 &= (- r_h^4 V_1^2 + P^2) g_1 - 2 r_h^2 P V_1 g_2 - r_h^4 V_1^2 g_3\,,\\
	\Phi_2 &= r_h^2 (g_1 + g_3) V_1 + P g_2 \,,\qquad \\
	\gamma_{g_1 g_3} &= 1 + g_1 + g_3\,.\label{defgamma}
}
The coupling functions $g_1$, $g_2$, $g_3$ are evaluated on the horizon with $(\phi,X)|_{r = r_h} = (\phi_0,0)$. We note that the authors of Ref.~\cite{Babichev:2017guv} have obtained hairy BH solutions with non-zero $\phi$ and $X$ at the horizon in the shift symmetric beyond Horndeski theories. In such theories, by virtue of the shift symmetry, the vanishing Noether current gives rise to the solutions with $\phi'$ divergent at the horizon while the kinetic term $X$ remains finite there. In the scope of this paper, we focus on the theories described by Eq.~\eqref{eq:exEMS_action} in which the shift symmetry of scalar field does not hold due to the existence of $\phi$-dependence in the coupling functions $g_1$, $g_2$, and $g_3$. Hence, we focus on the solutions endowed with scalar field, which is regular at the horizon.
As seen in Eq.~(\ref{eq:sol_horizon_f2}), the couplings give rise to corrections to the metric components at the order of $(r - r_h)^2$. 
When the coupling does not depend on the field $\phi$ itself, the coefficient $\phi_1$ 
in Eq.~\eqref{eq:sol_horizon_phi1} identically vanishes which results in the elimination of 
$\phi_2$ given in Eq.~\eqref{eq:sol_horizon_end}. 
In such a case, we can numerically confirm that the scalar field settles down to be constant 
$\phi = \phi_0$ in all the region outside the horizon which implies the absence of scalar hair. 
As a consequence, we find that the $\phi$-dependence in the functions $g_i(\phi,X)$ is 
essential for achieving hairy BH solutions.
%
\subsection{Asymptotically flat solutions}
\label{sec3b}
We derive asymptotically flat solutions satisfying $f, h \rightarrow 1$, 
$V \rightarrow V_{\infty}$, and $\phi \rightarrow \phi_{\infty}$ as $r \rightarrow \infty$,
where $V_{\infty}$ and $\phi_{\infty}$ are constants.
To obtain the solutions at spatial infinity, we expand $f$, $h$, $V$, $\phi$ 
as the power series of $1/r$, as
\be
	f = 1 + \sum_{i=1}^{\infty} {\tilde{f}_i \over r^i}\,,\qquad
	h = 1 + \sum_{i=1}^{\infty} {\tilde{h}_i \over r^i}\,,\qquad
	V = V_{\infty} + \sum_{i=1}^{\infty} {\tilde{V}_i \over r^i}\,,\qquad
	\phi = \phi_{\infty} + \sum_{i=1}^{\infty} {\tilde{\phi}_i \over r^i}\,.
	\label{eq:expansion_inf}
\ee
We substitute Eq.~(\ref{eq:expansion_inf}) into the background equations given in 
Eqs.~(\ref{eq:BG})-(\ref{eq:BG_end}) and expand them for the large $r$.  
Solving the resultant equations order by order in terms of inverse power of $r$, 
we find the following iterative solutions,
\myalign{
	f &= 1 - {2M \over r} + {\gamma_{g_1g_3} Q^2 + (1 + g_1) P^2 \over 2\Mpl^2 r^2}
			 + {Q_s \left[ M Q_s - (g_{1,\phi}+g_{3,\phi}) Q^2 + 2 g_{2,\phi} PQ + g_{1,\phi} P^2 \right] 
			 	 	\over 6\Mpl^2 r^3} 
			 + \mathcal{O}\left({1\over r^4}\right),
			 \label{eq:sol_inf}
\\
	h &= 1 - {2M \over r} + {\gamma_{g_1g_3} Q^2 + Q_s^2 + (1 + g_1) P^2 \over 2\Mpl^2 r^2}
			 + {Q_s \left[ M Q_s - (g_{1,\phi}+g_{3,\phi}) Q^2 + 2 g_{2,\phi} PQ + g_{1,\phi} P^2 \right] 
			 	 	\over 2\Mpl^2 r^3} 
			 + \mathcal{O}\left({1\over r^4}\right),
\\
	V &= V_{\infty} + {Q \over r} 
			 - {\left[ (g_{1,\phi}+g_{3,\phi}) Q - P g_{2,\phi} \right] Q_s \over 2 \gamma_{g_1g_3} r^2}
			 + \mathcal{O}\left({1\over r^3}\right),
\\
	\phi &= \phi_{\infty} + {Q_s \over r}
					+ {1 \over r^2} \left( 
														M Q_s 
														- {(g_{1,\phi}+g_{3,\phi}) Q^2 - 2 g_{2,\phi}PQ - g_{1,\phi}P^2 \over 4}
													\right)
					+ \mathcal{O}\left({1\over r^3}\right)\,,
				\label{eq:sol_inf_end}
}
where the coupling functions are evaluated at spatial infinity, i.e.,
$\left.g_i(\phi, X)\right|_{r\rightarrow \infty} = g_i(\phi_{\infty}, 0)$.
Among these couplings, $g_1$ and $g_3$ start to appear in the metric components 
$f$ and $h$ at the order of $1/r^2$, while $g_2$ first appears at the order of $1/r^3$. 
We also find that the scalar charge $Q_s$ appears in the metric component $h$ 
at the order of $1/r^2$ while it does not appear in $f$ at the same order.
Remembering that $f=h$ holds in the RN solution, it shows that the existence of scalar charge $Q_s$ 
triggers the deviation of solutions from the RN metric far away from the horizon. 
If the coupling has no $\phi$-dependence, Eq.~(\ref{eq:sol_inf_end}) reads 
%
$\phi = \phi_{\infty} + ({Q_s}/{r}) ( 1 + {M}/{r} )$. 
%
Although the scalar field seems to be nontrivial, this scalar charge $Q_s$ should 
vanish since the scalar profile is constant in all regions outside the horizon 
from the discussion in the last of the previous subsection.
Then, the metric components reduce to RN solution under the redefinition of $Q$ and $P$.
As a consequence, we conclude that the $\phi$-dependence of the functions $g_i$ 
generates the nonzero scalar charge $Q_s$ and the modification to the RN solution. 
%
%
\section{Concrete models and numerical analyses}
\label{sec4}
In this section, we study numerical analyses in concrete models with fixed coupling functions. 
We numerically solve the background equations given in Eqs.~(\ref{eq:BG})-(\ref{eq:BG_end}) 
outside the horizon by using the analytic solutions (\ref{eq:sol_horizon})-(\ref{eq:sol_horizon_end}) 
as boundary conditions around $r = r_h$.
To analyze the fundamental effects of each coupling, we set the functions $g_i (\phi, X)$ 
to be linear in terms of $\phi$ and $X$ as
\be
	g_i(\phi,X)\,  = \frac{c_i \phi}{\Mpl} \left( 1 + \frac{d_i X}{\Mpl^4} \right) \qquad
	{\rm with}\hspace{5pt} i \in \{1, 2, 3\}\,,
	\label{eq:concrete_models}
\ee
where $c_i$ and $d_i$ are coupling constants. 
Remembering that our interest is in the effect of derivative couplings on BH solutions, 
the coupling function $g_3$ characterizing the derivative interaction $Y$ plays a key role. 
The derivative interactions appearing through the $X$-dependence in $g_i$ are considered 
to contribute only as corrections to the $\phi$-dependence since the scalar hair is generated 
by the $\phi$-dependence in $g_i$ as we discussed in Sec.~\ref{sec3}. 
%
%
\subsection{Hairy BH solution arising from the derivative interaction $\bar{g}_3Y$}
\label{sec4a}
\begin{figure}[t]
	\centering
	\includegraphics[width=0.5\hsize]{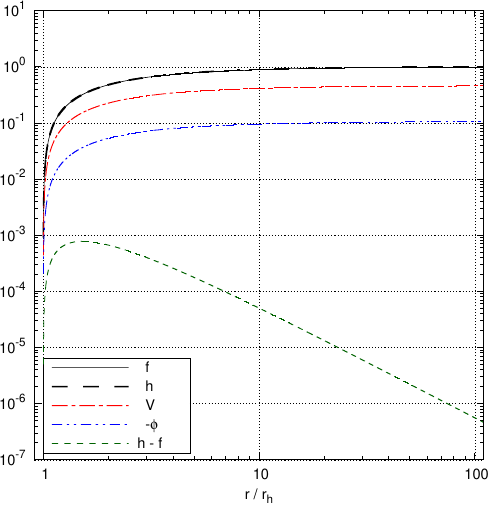}	
	\caption{Numerical solutions of $f, h, V, \phi, h-f$ outside the horizon for the coupling 
	$c_3 = 1$, with vanishing $c_1$, $c_2$, and $d_i$. The boundary conditions are chosen to 
	satisfy Eqs.~(\ref{eq:sol_horizon})-(\ref{eq:sol_horizon_end}) with $\mu = 0.1$, $P = 0.1 \Mpl r_h$, 
	$\phi_0 = -1.0\times10^{-4}\Mpl$ at $r = 1.001 r_h$.}
	\label{fig:c3_1_d3_0}
\end{figure}

For the scalar and the axionic-type couplings of the forms 
$g_1F$ and $g_2\tilde{F}$, respectively, 
it has shown that the charged BH solutions with scalar hair exist 
\cite{Gibbons:1987ps, Garfinkle:1990qj, Lee:1991jw, Boskovic:2018lkj, Astefanesei:2019pfq, Fernandes:2019kmh}. 
This subsection aims to examine whether the new BH solutions arise
from the derivative interaction, $\bar{g}_3Y$, in intermediate scales. 
For this purpose, we concentrate on the case $c_3=1$ with vanishing $c_1$, $c_2$, and $d_i$. 
We solve the field equations given in Eqs.~(\ref{eq:BG})-(\ref{eq:BG_end}) numerically 
with the model parameters $\mu = 0.1$ and $P = 0.1 \Mpl r_h$.
The analytic solutions around the horizon given in Eqs.~(\ref{eq:sol_horizon})-(\ref{eq:sol_horizon_end}) 
are used as boundary conditions with $\phi_0 = -1.0\times10^{-4}\Mpl$ at $r = 1.001 r_h$ 
\footnote{The boundary conditions and model parameters are chosen so that
(i) a condition for the couplings to work as corrections to the RN solution, 
(ii) a condition for avoiding discontinuity, 
(iii) a condition for the existence of horizon, 
are satisfied. The concrete expressions of these conditions will be discussed 
in Sec.~\ref{sec4b}.}.
We present the result in Fig.~\ref{fig:c3_1_d3_0}. 
It shows that the two asymptotic solutions, in the vicinity of the horizon and at the spatial infinity, 
connect to each other without any singular behavior. 
This is a new type of BH solution with scalar hair arising from the derivative interaction $\bar{g}_3Y$. 
In this case, the combination $h - f$ characterizing deviation of solutions from the RN metric 
takes the maximum value of order $h - f\sim\mathcal{O}(10^{-3})$ around the horizon, 
and decreases for larger $r$.
We note that, although $f$ generally converges to the value being different from $1$ at the large distances 
in numerical calculations while the metric $h$ settles down to $1$ there, we can normalize $f$ to satisfy 
$f(r \rightarrow \infty)=1$ by using the indefiniteness of $f$ in terms of time-rescaling.  
Indeed, in our numerical calculation, the normalization factor is determined by solving 
the background equations up to $r = 10^{16} r_h$. In the next subsection, we investigate 
the characteristics of this solution compared to those arising from $g_1F$ and $g_2\tilde{F}$. 
%
%
\subsection{Large coupling limit of each interaction and possible parameter space}
\label{sec4b}
As we discussed in Sec.~\ref{sec2}, Eq.~(\ref{eq:bg_quantities}) shows that 
the coupling $g_2\tilde{F}$ identically vanishes and the effect of $\bar{g}_3Y$ 
will be degenerated with one in $g_1F$ in the absence of magnetic charge $P$. 
In this subsection, we focus on the case of $P\neq0$ and examine how we can 
distinguish these three types of interactions considering the large coupling limit. 

The combination $|h - f|$, which characterizes deviation from the RN solution, 
can be expressed analytically around the horizon from Eq.~(\ref{eq:sol_horizon_f2}) as 
\be
	|h - f| = \left|\frac{\Phi_{1,\phi}^2}
{2 \Mpl^2 r_h^2 \, (1-\mu)(2r_h^4-\Phi_{1,X})}\right|\,
(r - r_h)^2 + \mathcal{O}\left( (r - r_h)^3 \right)\,.
				\label{eq:metric_difference}
\ee
To distinguish how the deviation between two metric components, $f$ and $h$, are affected 
by each coupling, we consider a large coupling limit characterized by $|c_i|\gg1$.
Expanding the subtraction of metric components (\ref{eq:metric_difference}) in a series of large $|c_i|$ 
for the three cases (i) $|c_1|\gg\{|c_2|,\,|c_3|\}$, (ii) $|c_2|\gg\{|c_1|,\,|c_3|\}$, and (iii)
$|c_3|\gg\{|c_1|,\,|c_2|\}$, the leading-order contributions are given as 
\myalign{
\frac{|h - f|}{(r-r_h)^2} 
&\simeq 
\left|\frac{P^2 c_1 (2\mu M_{\rm pl}^3 r_h^2 - P^2 c_1 \phi_0)}
{(1-\mu)\phi_0 M_{\rm pl}^4 r_h^6} 
- \frac{\mu (\mu \Mpl^2 r_h^2 + 2 P^2)}{(1-\mu) \phi_0^2 r_h^4}\right|
+ \mathcal{O}\left(\frac{1}{c_1}\right)
&\left(|c_1|\gg\{|c_2|,\,|c_3|\}\right)\,,
\label{eq:large_c1}
\\
\frac{|h - f|}{(r-r_h)^2} 
&\simeq \frac{P^2 c_2^2 (2\mu M_{\rm pl}^2 r_h^2 - P^2)}
{(1-\mu)M_{\rm pl}^4 r_h^6} 
&\left(|c_2|\gg\{|c_1|,\,|c_3|\}\right)\,,
\label{eq:large_c2}
\\
\frac{|h - f|}{(r-r_h)^2} 
&\simeq \left|-\frac{(2\mu M_{\rm pl}^2 r_h^2 - P^2)^2}
{4(1-\mu)\phi_0^2 \Mpl^2 r_h^6} + 
\frac{(2\mu M_{\rm pl}^2 r_h^2 - P^2)^2}{2(1-\mu) c_3 \phi_0^3 M_{\rm pl} r_h^6}\right|
+ \mathcal{O}\left(\frac{1}{c_3^2}\right)
&\left(|c_3|\gg\{|c_1|,\,|c_2|\}\right)\,
\,,
\label{eq:large_c3}
}
respectively. 
Eq.~(\ref{eq:large_c1}) shows that, in the case (i) $|c_1|\gg\{|c_2|,\,|c_3|\}$, 
the contributions of quadratic term $c_1^2$ has opposite sign to that of linear term 
$c_1$ as long as $c_1\phi_0>0$. 
Then, there should be extreme value of $c_1$ at which the dominant contribution 
switches from the linear term to the quadratic term. 
In the case (ii) $|c_2|\gg\{|c_1|,\,|c_3|\}$, we find that the absolute value of 
quantity $h-f$ monotonically increases in proportion to $c_2^2$ from Eq.~\eqref{eq:large_c2}. 
The derivative interaction characterized by $c_3$ exhibits 
particular behavior at the large coupling limit. In the case (iii) 
$|c_3|\gg\{|c_1|,\,|c_2|\}$, Eq.~\eqref{eq:large_c3} shows that 
the quantity $h-f$ saturates in the large coupling limit $|c_3|\to\infty$. 
{\par}
We are going to confirm the above characteristic signatures of each coupling by investigating numerical solutions. 
In doing so, the model parameters $c_i$, $\mu$, $P$, $\phi_0$ cannot be freely chosen but 
constrained from the requirements listed below. 
\begin{enumerate}
	\item[(i)] {\it a condition for the couplings to work as corrections to the RN solution }:
	In Sec. \ref{sec3a}, we regarded the effect of couplings as corrections to the RN solution. 
	For the sake of consistency, the contributions originated from the couplings need to be sub-leading 
	in the metric components given in Eq.~(\ref{eq:sol_horizon_f2}) compared to those in the RN solution. 
	In other words, regarding the asymptotic solutions (\ref{eq:sol_horizon_f2}) with (\ref{eq:sol_horizon_phi1}),
	we demand the following condition, 
	\be
		\left| \frac{\Phi_{1,\phi}^2}{8(\Phi_{1,X} - 2r_h^4) \Mpl^2} \right|
			\ll
		\left| (1 - \mu) (2\mu - 1) \right|\,, 
		\label{eq:condition_(i)}
	\ee
	under which the correction terms remain subdominant compared to the RN solution. 
	\item[(ii)] {\it a condition for avoiding discontinuity }:
	The field equations (\ref{eq:BG_phi}) and (\ref{eq:BG_end}) contain the second derivatives $V''$ and $\phi''$.
	Denoting the coefficients of $\phi''$ and $V''$ in Eq.~(\ref{eq:BG_phi}) as $a$ and $b$, respectively, 
	they are represented as 
	\be
		a = \frac{J_{\phi}}{\phi'} - h \phi' \, \frac{\partial J_{\phi}}{\partial X}\,, \qquad 
		b = -h\phi'\left[r^2 \sqrt{\frac{h}{f}}\,\,(g_{1,X}+g_{3,X})\,V' + g_{2,X} P\right]\,.
	\ee
	We also denote the coefficients of $\phi''$ and $V''$ in Eq.~(\ref{eq:BG_end}) as $c$ and $d$, respectively, 
	which are given by 
	\be
		c = -h\phi' \,\frac{\partial J_A}{\partial X}\,,\qquad 
		d = r^2 \sqrt{\frac{h}{f}}\,\, (1+g_1+g_3)\,.
	\ee
	One need to solve Eqs.~(\ref{eq:BG_phi}) and (\ref{eq:BG_end}) with respect to $V''$ and $\phi''$ 
	when performing numerical calculation. In doing so, the combination $ad - bc$ appears in the denominator 
	of these solutions. We demand this combination not to cross zero, i.e., $ad-bc\neq0$, so that the quantities 
	$V''$ and $\phi''$ remain continuous during numerical calculation. In the absence of the couplings, 
	we obtain $a=-r^2\sqrt{fh}$, $b=0$, $c=0$, and $d=r^2\sqrt{h/f}$, which lead $ad-bc=-r^4h<0$. 
	In order to realize this RN limit when $g_i \rightarrow 0$, we additionally require,
	\be
		ad - bc < 0\,.
		\label{eq:condition_(ii)}
	\ee
	We must choose the coupling constants $c_i$, $d_i$ as well as the model parameters $\mu$, $P$, $\phi_0$ 
	so as to satisfy the condition \eqref{eq:condition_(ii)} for any $r \, (r > r_h)$.
	In the absence of the kinetic couplings described by the $X$-dependence in the coupling functions, i.e., $g_{i,X} = 0$, 
	the coefficients $b$ and $c$ identically vanish while $a$ reduces to $a=-r^2\sqrt{fh}$. Then, the condition 
	\eqref{eq:condition_(ii)} simply reduce to $1+g_1+g_3>0$. This corresponds to the positivity of the quantity 
	$\gamma_{g_1 g_3}$ defined in Eq. (\ref{defgamma}). We note that the coupling $g_2$ is not constrained in such a case.
	\item[(iii)] {\it a condition for the existence of horizon }:
	For the existence of a real solution in the vicinity of the horizon, we require the positivity of 
	the term inside the square root of $V_1$ given in Eq.~(\ref{eq:sol_horizon}).
	This requirement leads to the following inequality
	\be
		\gamma_{g_1g_3} \left[2\mu \Mpl^2 r_h^2 - (1 + g_1)P^2\right] \ge 0\,.
		\label{eq:condition_(iii)}
	\ee
	This condition recovers the upper bound on electric and magnetic charge
	in the RN solutions if $g_1 = g_3 = 0$.
	The coupling $g_2$ itself is free from this requirement.
\end{enumerate}
We note that these three conditions are derived for the general coupling $g_i(\phi,X)$. Thus, we can apply these 
conditions even in the presence of the derivative couplings characterized by $d_i$ in Eq.~\eqref{eq:concrete_models}. 
\par 
\begin{figure}[b]
	\begin{minipage}{0.49\hsize}
		\centering
		\includegraphics[width=0.8\hsize]{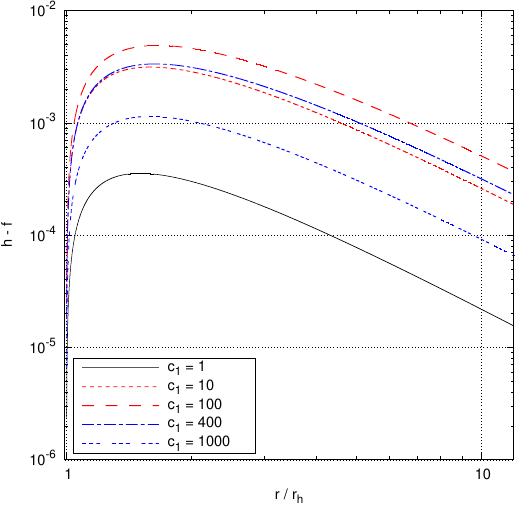}
		\caption{Numerical solutions of $h-f$ for different coupling constants, $c_1 \in \{ 1, 10, 100, 400, 1000 \}$. The model parameters are chosen as $\mu=0.1, P = 0.01\Mpl r_h$. The boundary condition is determined to satisfy Eqs.~(\ref{eq:sol_horizon})-(\ref{eq:sol_horizon_end}) at $r = 1.001 r_h$ with $\phi_0 = 0.5\Mpl$.}
		\label{fig:large_c1}
	\end{minipage}
	\hfill
	\begin{minipage}{0.49\hsize}
		\centering
		\includegraphics[width=0.8\hsize]{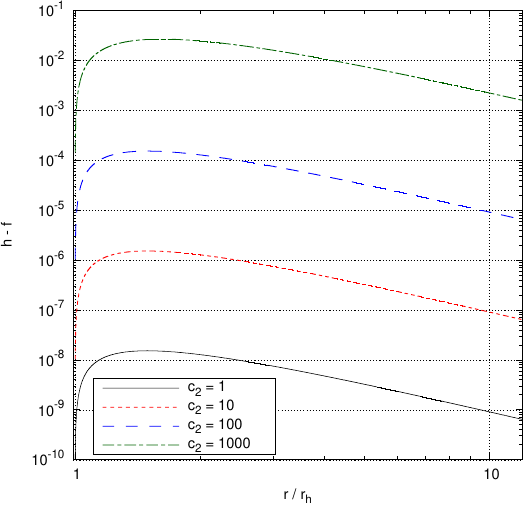}
		\caption{Comparing numerical solutions of $h-f$ between different coupling constants, $c_2 \in \{ 1, 10, 100, 1000 \}$. The model parameters are $\mu=0.1, P = 0.001\Mpl r_h$. The boundary condition is decided to satisfy Eqs.~(\ref{eq:sol_horizon})-(\ref{eq:sol_horizon_end}) at $r = 1.001 r_h$ with $\phi_0 = 0.1\Mpl$.}
		\label{fig:large_c2}
	\end{minipage}
\end{figure}
\begin{figure}[t]
		\centering
		\includegraphics[width=0.41\hsize]{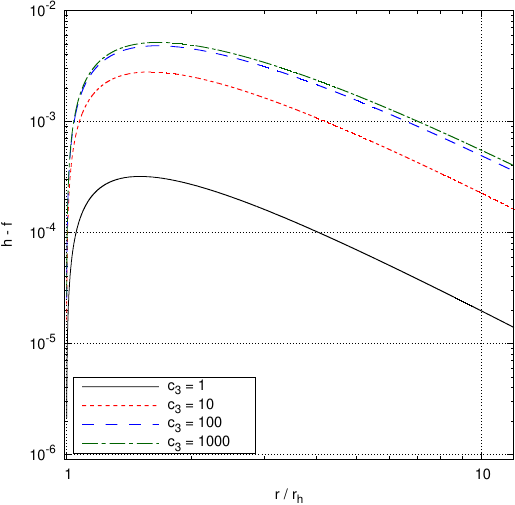}
		\caption{Numerical solutions of $h-f$ for different coupling constants, $c_3 \in \{ 1, 10, 100, 1000 \}$. The model parameters are chosen as $\mu=0.1, P = 0.1\Mpl r_h$. The boundary condition is determined to satisfy Eqs.~(\ref{eq:sol_horizon})-(\ref{eq:sol_horizon_end}) at $r = 1.001 r_h$ with $\phi_0 = 0.5\Mpl$.}
		\label{fig:large_c3}
\end{figure}
Figs.~\ref{fig:large_c1}-\ref{fig:large_c3} show the behavior 
of $|h - f|$ for the several values of $c_1$, $c_2$, or $c_3$.
The numerical results support the analytic solutions given in Eqs.~(\ref{eq:large_c1})-(\ref{eq:large_c3}).
From Eq.~(\ref{eq:large_c1}), there is a threshold at which the dominant contribution 
switches from the linear term to the quadratic term. 
%
For increasing $c_1$, our numerical result in Fig.~\ref{fig:large_c1} shows that the metric difference 
grows first, but it starts to be suppressed around $c_1\gtrsim{\cal O}(100)$.
In terms of large $c_2$, $|h - f|$ increases monotonically in Fig.~\ref{fig:large_c2}. 
This result shows good agreement with Eq.~(\ref{eq:large_c2}). 
However, $c_2$ cannot be large without the limit since the upper bound (\ref{eq:condition_(i)}) exists.
For a too large $c_2$ to violate the condition (\ref{eq:condition_(i)}) with fixed $P$, 
numerical integration tends to be unstable. 
Indeed, we numerically confirmed that the singular behavior starts to appear around 
$c_2 = 1870$, for the model parameters $\mu = 0.1$, $P = 0.001 \Mpl r_h$, $\phi_0 = 0.1 \Mpl$.
The violation of the condition (\ref{eq:condition_(i)}) implies that the resultant spacetime 
significantly deviates from RN spacetime. Since the asymptotic solutions 
(\ref{eq:sol_horizon})-(\ref{eq:sol_horizon_end}) around the horizon are obtained under 
the assumption that the couplings behave as corrections to the RN solution, they cannot 
approximate the solutions violating the condition (\ref{eq:condition_(i)}).
We note that the large $c_2$ limit has been discussed in Ref.~\cite{Lee:1991jw} 
under the assumption of the weak gravitational field. 
It has been analytically shown that the solutions around the horizon ($r \gtrsim r_h$) reduce to 
the RN metric possessing only a magnetic charge in such a case.  
In contrast to the previous study, we do not assume the weak field approximation and 
find that the solutions can deviate from the RN metric in large $c_2$ but singular behavior occurs 
at some point. 
When taking large $c_3$, Eq.~(\ref{eq:large_c3}) implies that the metric difference converges 
to finite values.
We numerically confirmed that this saturation realizes around $c_3\gtrsim{\cal O}(100)$ as shown in Fig.~\ref{fig:large_c3}. 
Hence, we can distinguish the three couplings by $|h - f|$ in the large coupling limit.
%
\subsection{Effects of $X$-dependence in $g_i(\phi,X)$ on scalar hair}
\label{sec4c}
\begin{figure}[H]
		\centering
		\includegraphics[width=0.5\hsize]{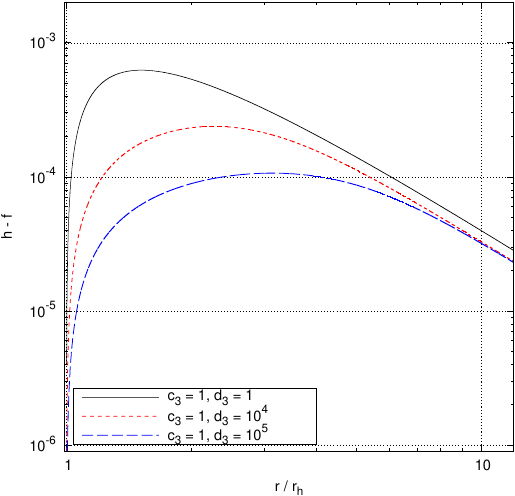}
		\caption{Numerical solutions of $h - f$ for different coupling constants, 
		$d_3/(r_h\Mpl)^2 \in \{ 1, 10^4, 10^5 \}$ with $c_3 = 1$. The boundary conditions are chosen to satisfy Eqs.~(\ref{eq:sol_horizon})-(\ref{eq:sol_horizon_end}) at $r = 1.001r_h$ with $\mu = 0.1, P = 0.1 \Mpl r_h, \phi_0 = 0.1 \Mpl$.}
		\label{fig:large_d3}
\end{figure}

In this subsection, we investigate the effects of the derivative interaction through the 
$X$-dependence in the coupling functions $g_i(\phi,X)$. 
As we have discussed in Sec.~\ref{sec3}, this type of coupling itself does not give rise to scalar hair. 
However, it can affect the hairy BH solutions originated from the $\phi$-dependence 
that we investigated in Sec.~\ref{sec4a} and \ref{sec4b}. 
In order to reveal this effect in detail, we focus on the quantity $h-f$ representing modification to the RN solutions 
and examine how its behavior changes depending on $d_i$ which describes the strength of 
kinetic interactions between the scalar field and $U(1)$ gauge field.
Although the kinetic term $X$ vanishes at the horizon as a result of the regularity of the scalar field, 
the kinetic coupling $d_i$ can affect the solutions around the horizon, $f_2$ and $h_2$, given in 
Eq.~(\ref{eq:sol_horizon_f2}) with (\ref{eq:sol_horizon_phi1}) through the coupling term $\Phi_{1,X}$.
Since the term $\Phi_{1,X}$ appears only in the denominator of these solutions, 
the deviation from the RN solution given in Eq.~(\ref{eq:metric_difference}) should be 
suppressed when the $X$-dependence in $\Phi_1$ is significantly large compared to the $\phi$-dependence. 
To clarify the difference among the effects of $d_1$, $d_2$, and $d_3$ on $h-f$,
we take a large coupling limit with respect to each $|d_i|$ in the same manner as in Sec.~\ref{sec4b}.
For the three cases (i) $|d_1|\gg\{|d_2|,\,|d_3|\},$ (ii) $|d_2|\gg\{|d_1|,\,|d_3|\}$, (iii) $|d_3|\gg\{|d_1|,\,|d_2|\}$,
we expand the metric difference (\ref{eq:metric_difference}) and find the leading-order contributions 
described by
\myalign{
\frac{|h - f|}{(r-r_h)^2} 
&\simeq 
\left| 
	- \frac{\left[ \mu \Mpl^3 r_h^2 - P^2 (c_1\phi_0+\Mpl) \right] c_1\Mpl}
			 {(1-\mu)(c_1\phi_0+\Mpl)\,d_1\phi_0 r_h^2} 
	+ \frac{\Mpl^6 r_h^2}{(1-\mu)\,d_1^2 \phi_0^2}
\right| + \mathcal{O}\left(\frac{1}{d_1^3}\right)
&\left(|d_1|\gg\{|d_2|,\,|d_3|\}\right)\,,
\label{eq:large_d1}
\\
\frac{|h - f|}{(r-r_h)^2} 
&\simeq
\left|
	- \frac{P c_2 \Mpl \sqrt{2\mu\Mpl^2r_h^2 - P^2}}{(1-\mu)\,d_2 \phi_0 r_h^2}
	+ \frac{\Mpl^6 r_h^2}{(1-\mu)\,d_2^2 \phi_0^2}
\right| + \mathcal{O}\left(\frac{1}{d_2^3}\right)
&\left(|d_2|\gg\{|d_1|,\,|d_3|\}\right)\,,
\label{eq:large_d2}
\\
\frac{|h - f|}{(r-r_h)^2} 
&\simeq
\left|
	- \frac{(2\mu \Mpl^2 r_h^2) c_3 \Mpl^2}{2(1-\mu)(c_3\phi_0+\Mpl)\,d_3\phi_0 r_h^2}
	+ \frac{\Mpl^6 r_h^2}{(1-\mu)\,d_3^2 \phi_0^2}
\right| + \mathcal{O}\left(\frac{1}{d_3^3}\right)
&\left(|d_3|\gg\{|d_1|,\,|d_2|\}\right)\,,
\label{eq:large_d3}
}
respectively.
Since each first term is proportional to $1/d_i$ in Eqs.~(\ref{eq:large_d1})-(\ref{eq:large_d3}), 
the qualitative behaviors of the metric difference for the cases (i), (ii), and (iii), are the same. 
Thus, we focus on the numerical solutions of $h - f$ only in terms of $d_3$ in the following.
In Fig.~\ref{fig:large_d3}, we compare the solutions of $h - f$ for different values of the constant $d_3$.
For the other model parameters, we adopt $c_3 = 1$, $\mu = 0.1$, $P = 0.1 \Mpl r_h$, $\phi_0 = 0.1 \Mpl$
with the boundary conditions (\ref{eq:sol_horizon})-(\ref{eq:sol_horizon_end}) at $r = 1.001 r_h$. 
We note that these parameters are chosen to be in the range of the conditions 
(\ref{eq:condition_(i)})-(\ref{eq:condition_(iii)}).
Fig.~\ref{fig:large_d3} shows the suppression of $h-f$ for increasing $d_3$ which is consistent with 
the aforementioned expectation for the effect of $d_i$. 
At a distance from the horizon, the kinetic term $X$ starts to evolve. This leads to a shift of 
the distance at the maximum point of $|h - f|$ to be larger for increasing $d_3$. 
We also confirmed that the effect of $d_i$ on the metric difference $h - f$ is the most crucial 
for $d_3$ among the three types of kinetic coupling $d_i$ as long as the value of the other 
parameters are similar. 
%
%
%
%
\section{Conclusions}
\label{sec5}
We studied BH solutions in extended Einstein-Maxwell-scalar theories
with magnetic charge and kinetic couplings.
In Sec. \ref{sec2}, we presented a subclass of
the $U(1)$ gauge-invariant scalar-vector-tensor theories
described by the quadratic interaction $f_2(\phi, X, F, \tilde{F}, Y)$.
To investigate the fundamental effects of the scalar-vector interactions on BH solutions,
we focused on a static and spherically symmetric spacetime whose line-element is 
given by Eq.~(\ref{eq:line-element}).
We introduced the scalar field $\phi$ and the vector potential $A_{\mu}$ under the ansatz (\ref{eq:ansatz}).
The $U(1)$ gauge field was set to have both electric and magnetic charges in order to 
resolve the degeneracy between the two types of derivative interactions, 
$\nabla_{\alpha}\phi \nabla^{\alpha}\phi\, F_{\mu\nu} F^{\mu\nu}$ and 
$\nabla_{\mu}\phi \nabla^{\nu}\phi\, F^{\mu\alpha} F_{\nu\alpha}$,
and to take account of the contribution from $F_{\mu\nu} \tilde{F}^{\mu\nu}$.
Introducing the magnetic charge breaks the spherical symmetry of the vector field configuration itself
while keeping the same symmetry of background equations.
Then, we focused on the theories described by Eq.~(\ref{eq:exEMS_action}) with the normalization (\ref{eq:g3_normalize}) 
so as to investigate the effects of the derivative coupling $Y$ and the $X$-dependence in the coupling functions 
on the BH solutions.
For the aforementioned configurations, the background equations were derived in Eqs.~(\ref{eq:BG})-(\ref{eq:BG_end}).
\par 
The aim of Sec.~\ref{sec3} is to find analytic solutions under proper approximations.
We focused on the two asymptotic regions, in the vicinity of the horizon, $r \simeq r_h$, and the spatial infinity, $r \gg r_h$.
In Sec.~\ref{sec3a}, we expanded the equations of motion with respect to $r - r_h$, 
demanding regularity of variables at $r = r_h$ as in Eq.~(\ref{eq:expansion_horizon}). 
In the vicinity of the horizon, we obtained the expansion coefficients as in Eqs.~(\ref{eq:sol_horizon})-(\ref{eq:sol_horizon_end}).
The deviation of metric components from the RN solution starts to arise at the order of $(r - r_h)^2$.
In the absence of $\phi$-dependence in the coupling functions,
the metric settles down to the RN solution, and the scalar profile becomes trivial.
This shows that the crucial source of scalar hair is the $\phi$-dependence in the coupling functions.
The kinetic couplings characterized by the $X$-dependence in the coupling functions 
can contribute as modifications to the existing hairy solutions arising from the $\phi$-dependence of them.
In Sec.~\ref{sec3b}, in order to derive asymptotically flat solutions at large distances, 
the metric components, scalar, and vector fields are expanded in the power series of $1/r$ 
as in Eq.~\eqref{eq:expansion_inf}. 
We then solved the background equations order by order in terms of $1/r$, and determined 
the expansion coefficients in Eqs.~(\ref{eq:sol_inf})-(\ref{eq:sol_inf_end}), up to the order of $1/r^3$.
The metric components are corrected by $g_1$ and $g_3$ from the order of $1/r^2$,
while $g_2$ first arises in these solutions at the order of $1/r^3$.
We also found that the metric component $h$ includes the scalar charge $Q_s$ 
at the order of $1/r^2$, unlike the other component $f$.
Hence, the deviations from the RN solutions start to appear at the order of $1/r^2$ 
due to the existence of scalar charge $Q_s$.
When the couplings do not include $\phi$-dependence, the scalar charge $Q_s$ should vanish
to satisfy the consistency with the discussion in Sec.~\ref{sec3a}.
\par 
In Sec. \ref{sec4}, we fixed the functional form as in Eq.~(\ref{eq:concrete_models})
to perform numerical integrations of the field equations.
The subsection \ref{sec4a} focused on investigating the solutions
in the presence of derivative interaction $\bar{g}_3 Y$.
We used the asymptotic solutions given in Eqs.~(\ref{eq:sol_horizon})-(\ref{eq:sol_horizon_end}) 
near the horizon as boundary conditions for integration.
In Fig.~\ref{fig:c3_1_d3_0}, the numerical analysis showed that 
these analytic solutions around the horizon smoothly connected to 
asymptotically flat solutions given in Eqs.~(\ref{eq:sol_inf})-(\ref{eq:sol_inf_end}).
We confirmed the existence of new hairy BH solutions arising from the derivative coupling $\bar{g}_3 Y$.
In Sec.~\ref{sec4b}, we discussed the difference of hairy BH solutions 
originated from the three types of interactions $g_1 F$, $g_2 \tilde{F}$ and $\bar{g}_3 Y$.
To distinguish how the metric components are corrected by each coupling, 
we focused on the difference of metric components represented by $|h - f|$ 
in a large coupling limit characterized by $|c_i| \gg 1$.
The leading-order contributions were obtained in Eqs.~(\ref{eq:large_c1})-(\ref{eq:large_c3}).
We found that each solution shows distinguishable characteristic behavior in terms of $c_i$.
Before proceeding to numerical analysis, we search for the possible parameter space 
in the following way.
First, requiring the couplings to work as sub-leading corrections 
to the RN solution results in the condition (\ref{eq:condition_(i)}).
Second, we demand the condition (\ref{eq:condition_(ii)}) to avoid discontinuities of 
the second order derivatives of scalar field, $\phi''$, and vector field, $V''$.
Third, we require the existence of a real solution in the vicinity of the horizon, which 
results in the condition (\ref{eq:condition_(iii)}).
In Figs.~\ref{fig:large_c1}-\ref{fig:large_c3}, we compared the different behavior of $|h - f|$
for the several values of $c_1$, $c_2$ or $c_3$.
The numerical results are consistent with the analytic solutions given in Eqs.~(\ref{eq:large_c1})-(\ref{eq:large_c3}).
In Sec.~\ref{sec4c}, we studied the effects of the derivative interactions characterized by 
the $X$-dependence in the coupling functions~$g_i$.
When the $X$-dependence in $\Phi_1$ appearing in the solutions around 
the horizon (\ref{eq:sol_horizon})-(\ref{eq:sol_horizon_end})
dominates over the $\phi$-dependence, we found that 
the difference of the metric components, $|h - f|$, should be suppressed 
as in Eq.~(\ref{eq:metric_difference}) since $|h - f|$ includes $\Phi_{1,X}$ 
only in the denominator at the leading-order. 
For the three cases of the kinetic couplings $d_1$, $d_2$ and $d_3$, 
the quantity $|h - f|$ was expanded for large $|d_i|$ to give the leading-order 
contribution as in Eqs.~(\ref{eq:large_d1})-(\ref{eq:large_d3}).
Since the qualitative behavior for these three cases is the same, 
we focused on the numerical solutions of $h - f$ for different values of 
$d_3$ in Fig.~\ref{fig:large_d3}.
For increasing $d_3$, the numerical result shows the suppression of $h - f$ 
which is consistent with the analytic expression given in Eq.~(\ref{eq:large_d3}).
Fig.~\ref{fig:large_d3} also indicates the shift of the distance at the maximum point of $h - f$ 
to large $r$ in the region away from the horizon where the kinetic term $X$ grows.
\par 
In this paper, we focused on the hairy BH solutions
electrically and magnetically charged 
on the static and spherically symmetric background 
based on the quadratic interaction $f_2(\phi,X,F,\tilde{F},Y)$ 
in the $U(1)$ gauge-invariant SVT theories.
The hairy BH solutions originated from the cubic and the quartic interactions, $\mathcal{L}_3$ and $\mathcal{L}_4$, 
in the $U(1)$ gauge-invariant SVT theories were studied in Refs.~\cite{Heisenberg:2018vti} in the absence of magnetic charge. 
It would be of interest to investigate whether the presence of magnetic charge gives rise to 
a new type of hairy BH solution in the full $U(1)$ gauge-invariant SVT theories.
It is also of interest to step forward to BH perturbation analyses to examine the stability of our solutions. 
The BH perturbations within the theory including scalar-vector coupling were studied in 
Refs.~\cite{Heisenberg:2018mgr,Gannouji:2021oqz, Kase:2023kvq,Zhang:2024cbw} in the absence of magnetic charge. 
It is necessary to clarify how the existence of magnetic charge affects the dynamics of perturbations \cite{Nomura:2020tpc,Chen:2024hkm}
as well as stability conditions in the $U(1)$ gauge-invariant SVT theories.. 
%
%
%
\section*{Acknowledgements}

R. K. is supported by the Grant-in-Aid for Early-Career Scientists 
of the JSPS No.~20K14471 and the Grant-in-Aid for Scientific 
Research (C) of the JSPS No.~23K03421. 


\bibliography{grqc}
\bibliographystyle{apsrev4-2.bst}

\renewcommand{\appendix}{}
\appendix
\renewcommand{\thesection}{Appendix \Alph{section}}
\renewcommand{\thesubsection}{\Alph{section}.\arabic{subsection}}
\setcounter{section}{0}
\setcounter{subsection}{0}

\end{document}